\documentclass[conference]{IEEEtran}
\IEEEoverridecommandlockouts

\usepackage{amsmath,amssymb,amsfonts}
\usepackage{algorithmic}
\usepackage{textcomp}
\usepackage{xcolor}

\usepackage{cite}
\usepackage{times}
\usepackage{epsfig}
\usepackage{graphicx}
\usepackage{amsmath}
\usepackage{amssymb}
\usepackage{multirow}
\usepackage{longtable}
\usepackage{color}
\usepackage{rotating}
\usepackage{makecell}
\usepackage{array}
\usepackage{booktabs}
\usepackage{colortbl}
\usepackage{arydshln}

\usepackage{hyperref}
\usepackage{amsmath}        
\usepackage{bm}             
\usepackage{subfigure}
\usepackage{blindtext}

\usepackage{color}

\def\BibTeX{{\rm B\kern-.05em{\sc i\kern-.025em b}\kern-.08em
    T\kern-.1667em\lower.7ex\hbox{E}\kern-.125emX}}
\begin{document}

\title{Distribution-Aware Graph Representation Learning for Transient Stability Assessment of Power System

\author{\IEEEauthorblockN{Kaixuan Chen\textsuperscript{1}, Shunyu Liu\textsuperscript{1}, Na Yu\textsuperscript{1}, Rong Yan\textsuperscript{2}, Quan Zhang\textsuperscript{2}, Jie Song\textsuperscript{1}, Zunlei Feng\textsuperscript{1}, Mingli Song\textsuperscript{1,*}}
\IEEEauthorblockA{\textit{\textsuperscript{1} College of Computer Science and Technology, Zhejiang University, Hangzhou, China} \\
\textit{\textsuperscript{2} College of  Electrical Engineering, Zhejiang University, Hangzhou, China} \\
Email: \{chenkx, liushunyu, na\_yu, yanrong052, quanzzhang, sjie, zunleifeng, brooksong\}@zju.edu.cn  \\
}
\IEEEcompsocitemizethanks{
\IEEEcompsocthanksitem * corresponding author.}
}
\thanks{This work is funded by the National Key R\&D Program of China  (Grant No: 2018AAA0101503) and the Science and technology project of SGCC (State Grid Corporation of China): fundamental theory of human-in-the-loop hybrid-augmented intelligence for power grid dispatch and control.}
}

\maketitle

\begin{abstract}
The real-time transient stability assessment (TSA) plays a critical role in the secure operation of the power system. Although the classic numerical integration method, \textit{i.e.} time-domain simulation (TDS), has been widely used in industry practice, it is inevitably trapped in a high computational complexity due to the high latitude sophistication of the power system.
In this work, a data-driven power system estimation method is proposed to quickly predict the stability of the power system before TDS reaches the end of simulating time windows, which can reduce the average simulation time of stability assessment without loss of accuracy. 
As the topology of the power system is in the form of graph structure, graph neural network based representation learning is naturally suitable for learning the status of the power system. Motivated by observing the distribution information of crucial active power and reactive power on  the power system's bus nodes, we thus propose a distribution-aware learning~(DAL) module to explore an informative graph representation vector for describing the status of a power system. 
Then, TSA is re-defined as a binary classification task, and the stability of the system is determined directly from the resulting graph representation without numerical integration. Finally, we apply our method to the online TSA task.
The case studies on the IEEE 39-bus system and Polish 2383-bus system demonstrate the effectiveness of our proposed method. The code is available at
\url{https://github.com/kxchern/dkepool-tsa}\\
\end{abstract}


\section{Introduction}
Transient stability assessment (TSA)~\cite{gonzalez2021risk} is the necessary task for dynamic stability assessment of the power system, which involves the assessment of the power system's ability to remain synchronism after being subjected to some credible disturbances, such as a short circuit on a transmission line~\cite{kundur2004definition}. The transient instability of the power system may lead to catastrophic events, such as large-scale blackout and cascading failure. Therefore, maintaining transient is an essential requirement in power system control and operation.

\begin{figure}
  \vspace{-0.5cm}  
  \setlength{\abovecaptionskip}{0.2cm} 
  \setlength{\belowcaptionskip}{0.4cm} 
\centering
\hspace{-3mm}
\subfigure[Polish 2383-bus]{\includegraphics[width=4.65cm]{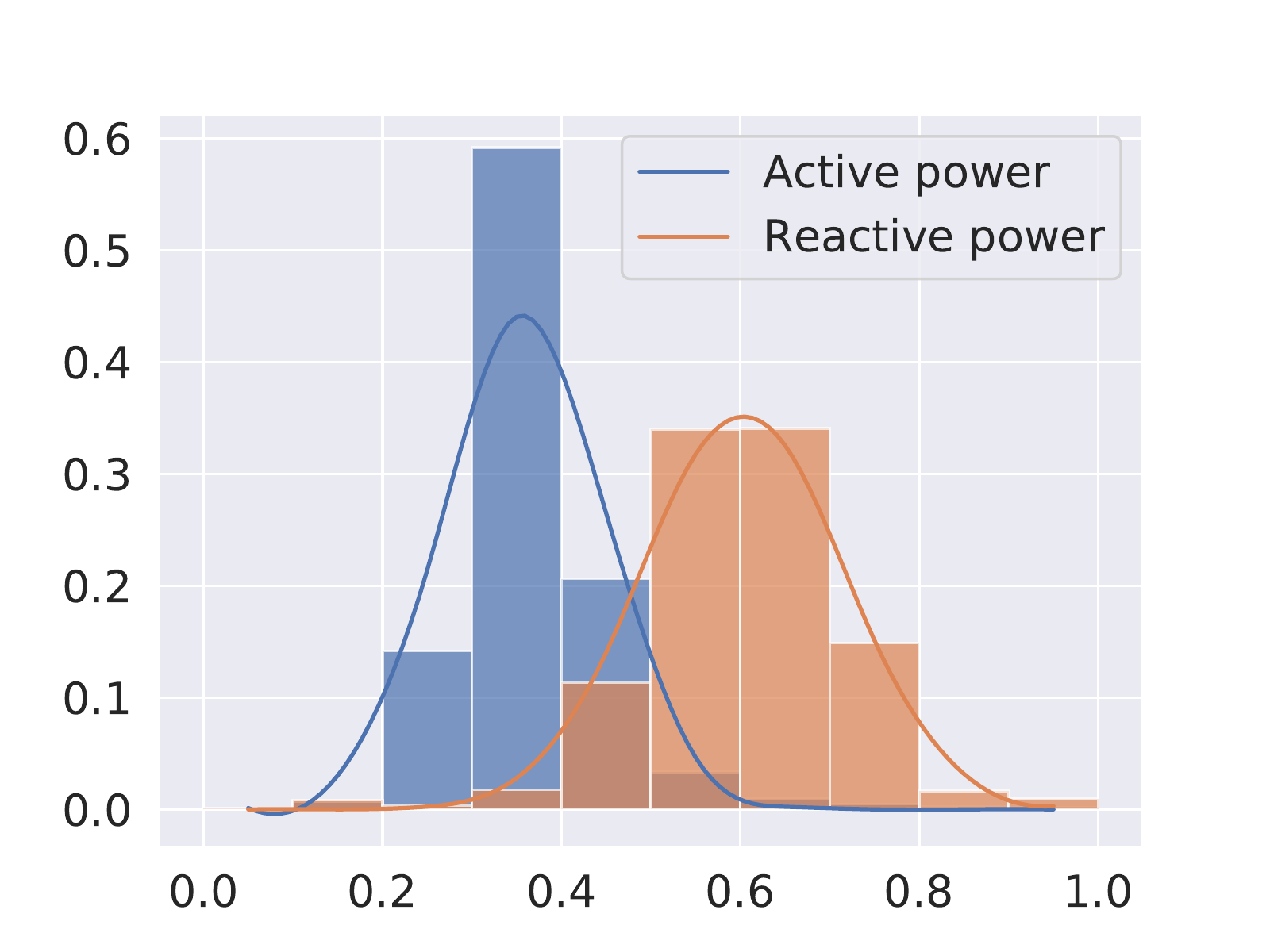}}
\hspace{-5mm}
\subfigure[IEEE 39-bus]{\includegraphics[width=4.65cm]{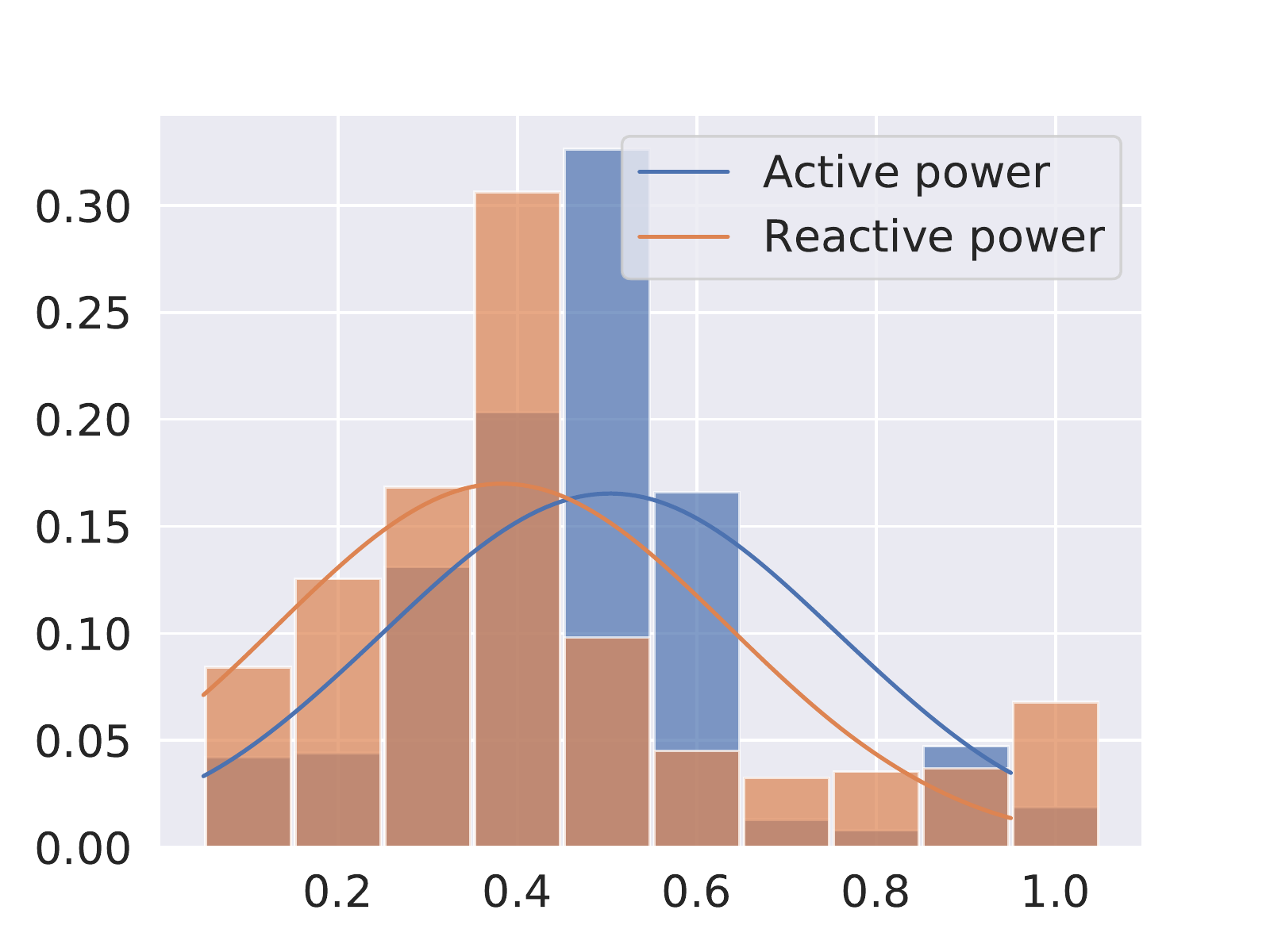}}
\hspace{-2mm}

\caption{
The illustration of distribution information in two classical power systems. (a) The distribution information of active and reactive power on bus nodes in the Polish 2383-bus system.  (b) The distribution information of active and reactive power on bus nodes in the IEEE 39-bus system.
}
\label{statistic}
\vspace{-0.5cm}  
\end{figure}

\begin{figure*}
  \vspace{-0.0cm}  
  \setlength{\abovecaptionskip}{-0.2cm} 
  \centering
  \includegraphics[scale=0.80]{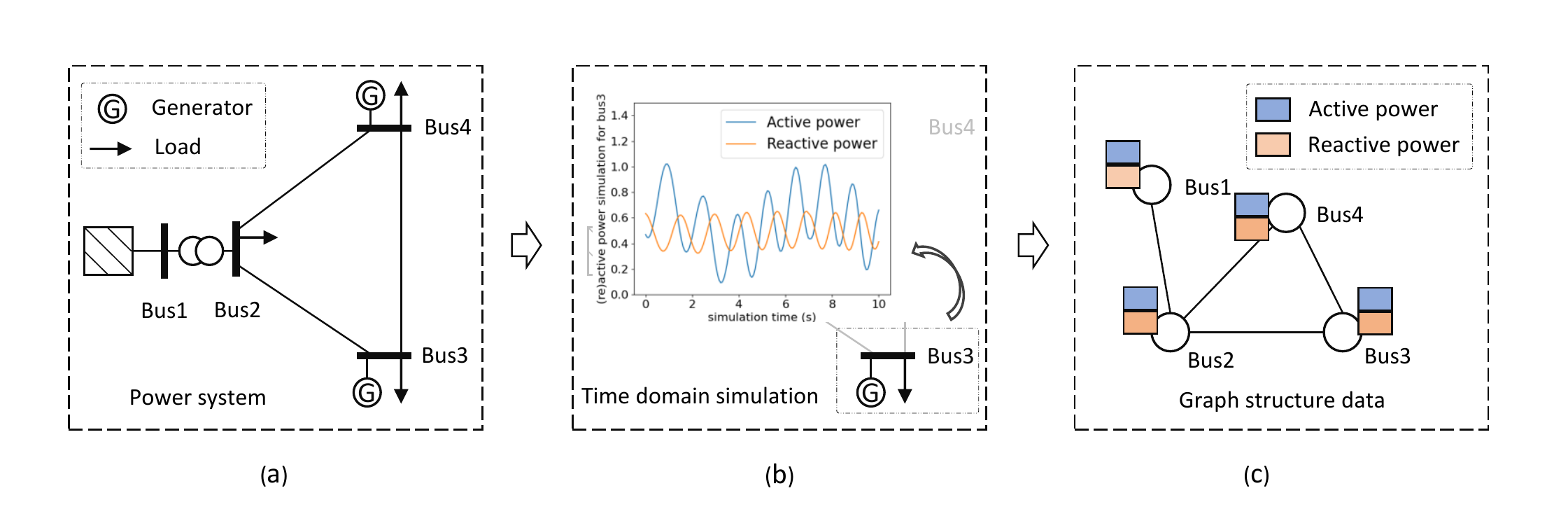}
  
  \caption{
An example of constructing power system status into graph structure data. (a) The topological structure of a simple power system with four bus nodes, two generators and three loads. (b) The classical time domain simulation method for analyzing the status of the power system. Each bus and its attached components is treated as an independent unit. (c) The graph structure data is modeled from the power system by considering the bus nodes as \emph{vertices} and transmission lines as \emph{edges}. The data processing method above that considers the topological structure information is crucial for downstream tasks that operate on the power system.
}
\label{fig:powersystem_2}
\vspace{-0.4cm}  
\end{figure*}

TSA can be used to predict the stability of the power system under continuous disturbance and use its evaluation decisions to trigger emergency control actions, such as generator trip and load shedding, which is very important to prevent unstable propagation. One of the most well-known and solid TSA methods is to evaluate the stability through time-domain simulation, which solves a set of high-dimensional nonlinear differential algebraic equations by an iterative method. However, TDS is computationally burdensome and requires accurate system modeling information, which will weaken its real-time performance. Therefore, an alternative framework to reduce TSA time without losing accuracy compared with TDS becomes a worthwhile topic.

With the rapid development of data-driven techniques, the status of the power system can be monitored online, which provides a way for real-time TSA. There have been a number of conventional machine learning methods suggested to speed up the TDS for TSA, such as decision trees (DT)~\cite{wang2020mvmo}, random forest (RF)~\cite{liu2021data}, support vector machine (SVM)~\cite{wang2020improved}. In data-driven TSA~\cite{yan2020data}, the primary evaluation process is to perform the designed intelligence model on the pre-prepared data usually obtained from simulation toolbox and historical records, and then apply the trained models online for TSA tasks with less computation effort. Furthermore, remarkable successes of the deep neural network have been achieved in the domain of signal processing, computer vision and natural language processing. In the domain of the power system, some deep models have been employed for system stability assessment and control, such as convolutional neural network (CNN)~\cite{yan2019fast} and deep reinforcement learning (DRL)~\cite{yoon2020winning}.

However, when it comes to power system TSA tasks, the performance of CNN will be weakened because of the ignore of the particular topology structure of power system data. As power system objects typically occur in the form of graph structure by considering the bus nodes and their connection information, it is reasonable and natural to consider using the graph neural network to learn the representation for describing the status of the power system. It is worth noting that the employment of the graph neural network to learn the status of the power system not only describes its topological structure but also matches the information transmission between bus nodes in the process of feature extraction.
Furthermore, the \emph{active} and \emph{reactive} power on bus nodes play a crucial role in predicting the transient stability of the power system, and \emph{their statistical information obeys the Gaussian distribution as shown in} Fig.~\ref{statistic}. Therefore, the most used readout operations like averaging and summing on the node feature will cause massive information loss, which may downgrade the performance of the resulting representation for the TSA task.

In this paper, we propose the \emph{distribution-aware learning} (DAL) module plugged into the graph neural network for graph representation learning with application to TSA task, which considers the above discussion about the topology of the power system and the distribution information of active and reactive power on the bus nodes. To this end, the representation learning for the status of the power system in this work is \emph{de facto} disassembled into two stages, \emph{i.e.}, \emph{structure-aware learning} and \emph{distribution-aware learning}. Structure-aware learning, powered by existing GNNs, follows a recursive neighborhood aggregation scheme to update node features where structure information is absorbed. Distribution-aware learning, on the other hand, omits node interconnections and focuses more on the distribution depicted by all the nodes. Our pooling goal is to learn a representation outlining the entire node distribution, which is used for the power system TSA task predictions.

The contributions of this work are summarized as following three folds:
\begin{itemize}
    \item We argue that learning the representation of the status of the power system for the TSA task should comprehensively consider information transmission among the bus nodes and distribution information of the active and reactive power on the bus nodes.

    \item We propose the \emph{distribution-aware learning} (DAL) module, an easy-to-use module plugged at the regular graph neural networks, to learn informative representations for describing the status of the power system. Then, We provide theoretical analysis to support why the proposed module can outline distribution information.

    \item We apply the resulting representation into the power system TSA task. The TSA decision can be delivered as quickly as possible while maintaining an acceptable accuracy, so emergency control actions can be timely and accurately triggered to avoid further blackout events.

\end{itemize}

The case studies on the IEEE 39-bus system and Polish 2383-bus system demonstrate the superior performance of our proposed module for the TSA task.  

\begin{figure*}
  \vspace{-0.0cm}  
  \centering
  \includegraphics[scale=0.575]{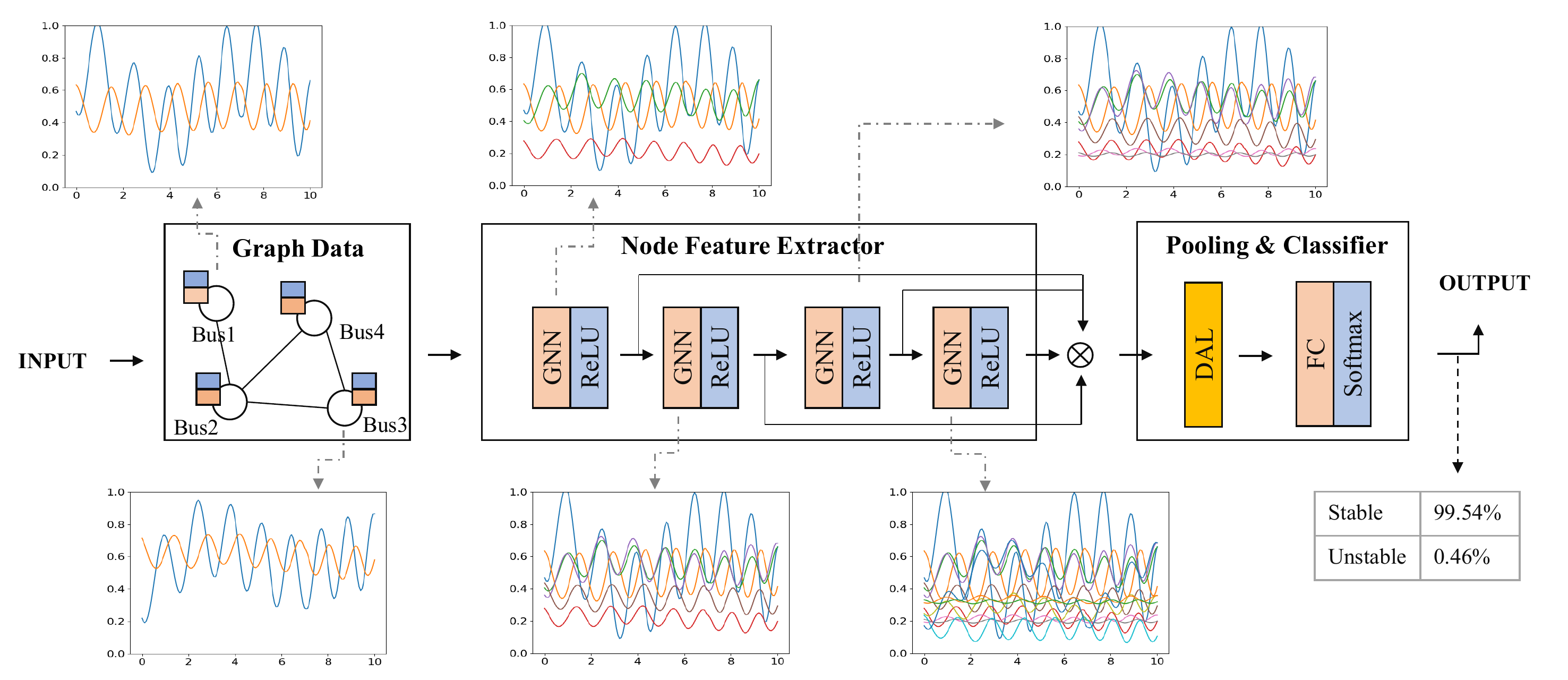}
  \caption{The illustration of our proposed framework for the TSA task. The GNN can learn the representation for each node, and our proposed module processes the node representation into an informative graph representation for describing the status of the power system. The FC layer is a linear classifier for the TSA problem.
}
\label{fig:framework}
\vspace{-0.4cm}  
\end{figure*}

\section{Problem Formulation}

The power system is a network of electrical components deployed to supply, transfer, and use electric power. It can be broadly divided into the generators that supply the power, the transmission system that carries the power from the generating centers to the load centers, and the distribution system that feeds the power to nearby homes and industries. The aim of this work is to assess the transient stability status of a power system while being subjected to some credible disturbances, which is of high importance to plan and initiate relevant corrective control actions. 

For a successful TSA scheme, the assessment methods should be sufficiently accurate and delivered as fast as possible while the power system is subjected to some credible disturbance. Although TDS is one of the most used methods for the TSA task, it is time consuming. Therefore, it is crucial to reduce the assessment  time and improve the TSA accuracy for online assessment. Next, we introduce how to assess the TSA problem by using the classical TDS method. Then, we model the power system data as graph structure data for constructing the data-driven dataset and re-define the TSA problem as a binary classification problem.

\subsection{TDS for TSA task}
As shown in Fig.~\ref{fig:powersystem_2} (a), the power system commonly includes bus, transmission line, generator and load, etc. In general, the electric power consumption and supply are achieved through the transmission lines between bus nodes, where generator and load are directly attached without transmission line. The power system model that includes lots of dynamic components can be described by a set of high-dimensional nonlinear differential algebraic equations. The classical TDS method can obtain the model results using the following numerical integration:
\begin{equation}
\label{differentialequations}
\begin{split}
\left\{\begin{matrix}
& \frac{d x}{d t}  = f(x,y), & \\ 
\\
& 0=g(x,y),  & 
\end{matrix}\right.
\end{split}
\end{equation} 
where $x$ denotes state variables that describe dynamics of the system in differential equations $f$, and $y$ denote operating variables in algebraic equations $g$.

The TDS can simulate the power system during a certain period of time after being subjected to some credible disturbances. As shown in Fig.~\ref{fig:powersystem_2} (b), we can obtain the value of each component during this period. At the end of the simulation, we assess whether the power system is stable by comparing synchronous generators' rotor angles. The transient stability index (TSI) can be calculated as follow:
\begin{equation}
\label{TSI}
\begin{split}
{\rm TSI} =  \frac{360-|\triangle \delta|_{max}}{360+|\triangle \delta|_{max}} \times 100,
\end{split}
\end{equation}
where $|\triangle \delta|_{max}$ is the absolute value of the maximum rotor
angle separation between any two generators. The status of power system $y$ can be assessed with TSI,
\begin{equation}
\label{labelstable}
\begin{split} y = 
\left\{\begin{matrix}
&1,   & {\rm while\; TSI>0}, \\ 
\\
&0,   & {\rm while\; TSI\leq 0,}\\
\end{matrix}\right.
\end{split}
\end{equation} 
where $y$ = 1 indicates the power system is stable, otherwise unstable.

\subsection{Data pre-processing and task re-defined}

We can compare the rotor angle values at the end of the time domain simulation, which simulates the power system's fluctuation by solving a set of high-dimensional nonlinear differential algebraic equations. Though it satisfies the requirements of reliability, accuracy and model adaptability, it is not suitable for online TSA because of the major drawback of the high computational burden. To this end, a number of direct methods that provide conservative and approximate assessment results have been proposed to speed up the TDS, such as Transient Energy Function (TEF)~\cite{bhui2016real} and Trajectory Convexity Concavity (TCC)~\cite{su2017study}. Moreover, the Machine Learning (ML) based methods have been applied for fast real-time TSA, which adopt the  conventional TDS data to construct the dataset for training the offline models and then perform the fast online TSA using the trained models. Among these methods,  DT~\cite{wang2020mvmo}, RF~\cite{liu2021data}, SVM~\cite{wang2020improved} and artificial
neural network (ANN)~\cite{guo2012multi}, have demonstrated their strengths in power system stability assessment. 

However, the above ML-based methods ignore the topology information of the power system that we argue is crucial to the resulting representation. Thus, we need to consider the power systems as graph structure data and adopt the graph neural network to learn the power system's status representation. In this paper, we use the TDS method to generate the input data for the data-driven model. Moreover, we treat each bus and its attached components as a node, and each transmission line as a line for message passing. For each bus, the key information includes active and reactive power as shown in Fig.~\ref{fig:powersystem_2} (b), which is crucial to determine the stability of the power system. Thus, we preserve the active and reactive power on bus nodes to construct the graph structure data. As shown in Fig.~\ref{fig:powersystem_2} (c), the power system data has been modeled as the graph structure data. Then, we label the data using Eq. (\ref{labelstable}).  For the data-driven model, we construct the power system status dataset  by looping the above operations multiple times. After that, we can re-define the TSA task as a binary classification problem to train the model.

\begin{figure*}
\vspace{-0.2cm}  
  \setlength{\abovecaptionskip}{0.1cm} 
  \centering
  \includegraphics[scale=0.49]{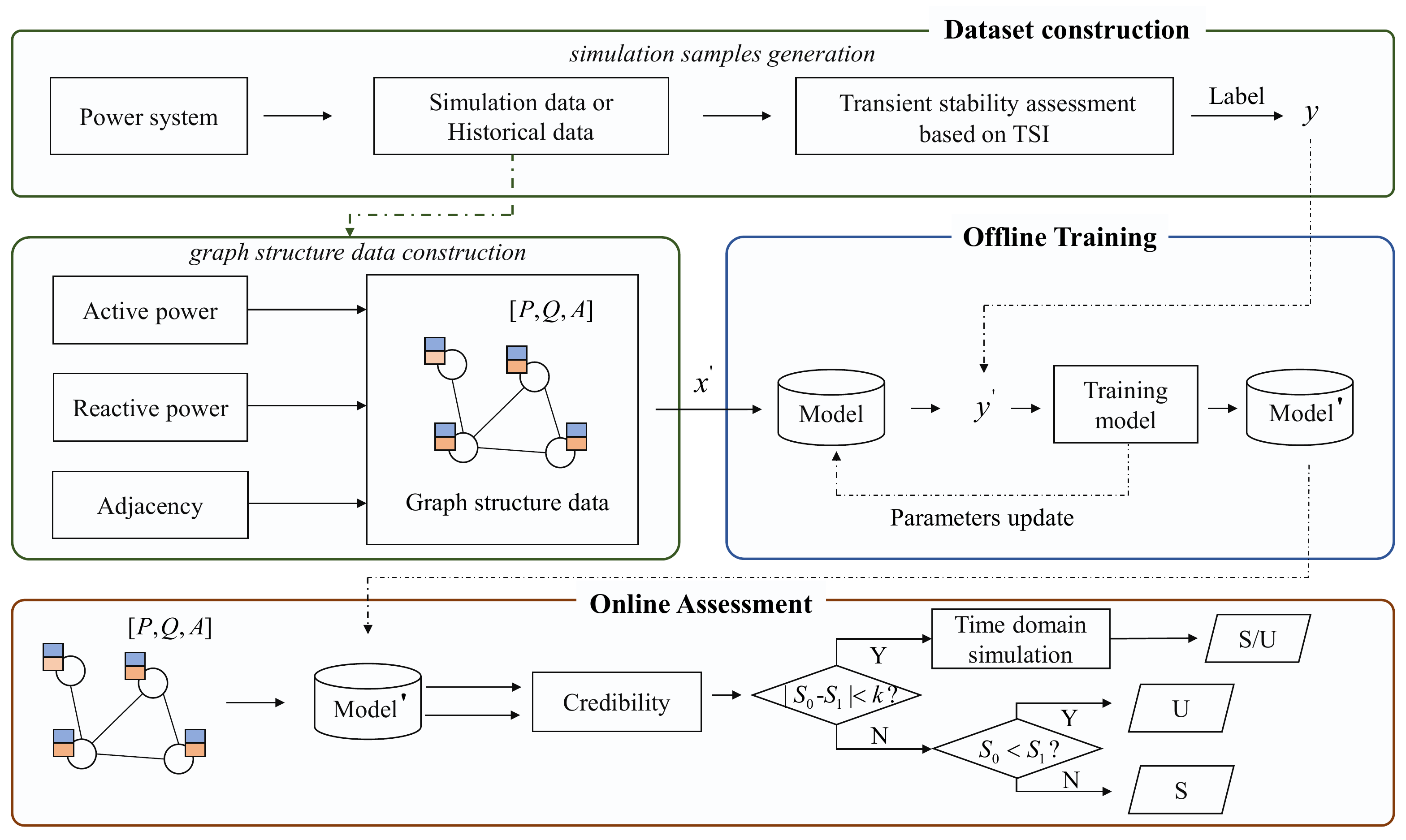}
  \caption{
	Algorithm flowchart of the online TSA. This task \emph{de facto} is disassembled into three stages, dataset construction, offline training and online assessment. The dataset construction includes \emph{simulation sample generation} and \emph{graph structure data construction} for the data-driven model. The offline training aims to train the model to select the best parameters. The trained model can be applied for the real-time TSA task at the online assessment stage.
}
\label{flowchart}
\vspace{-0.3cm}  
\end{figure*}

\section{TSA Task using distribution-aware graph representation learning}
In this section, we first present the definition of the graph neural network and its general paradigm. Then, we introduce our proposed \emph{distribution-aware learning} module to outline the distribution information in the form of vector representation. Finally, we apply the resulting graph representation for the TSA task.

\subsection{Graph neural networks}

A graph is a data structure consisting of two components, \emph{i.e.}, \emph{vertices}, and \emph{edges}. Formally, a graph consisting of $n$ nodes can be represented as $G=(\bm{A},\bm{X})$, where $\bm{A} \in \mathbb{R}^{n\times n}$ is adjacency matrix and $\bm{X}\in \mathbb{R}^{n\times d}$ is the node feature matrix. To efficiently aggregate node features with adjacency information, graph neural networks (GNNs)~\cite{cai2020multi,kipf2016semi,xu2018powerful,yang2020factorizable,hao2021walking,zheng2021learning} are developed to learn more powerful representations by considering topological structure information. Given a graph consisting of $n$ nodes, GNNs generally follow a message-passing architecture:
\begin{equation}
\label{massagepassing}
\begin{split}
 \bm{H}^{(k)}=M(\bm{A},\bm{H}^{(k-1)};{\theta}^{(k)}),
\end{split}
\end{equation} 
where $\bm{H}^{(k)}$ is the node features of the $k$-th layer and $M$ is the message propagation function. The trainable
parameters are denoted by $\theta^{(k)}$ and the adjacency matrix by $\bm{A}$. $\bm{H}^{(0)}$ is initialized as $\bm{H}^{(0)}=\bm{X}$. 

The propagation function $M$ can be implemented in various manners~\cite{cai2020multi,kipf2016semi,xu2018powerful,yang2020factorizable,hao2021walking,zheng2021learning}.  The recursive neighborhood aggregation scheme enables the node features to absorb the structure information. To this end, graph topology structure information has been learned into node features via GNNs. The most frequently used pooling operations like averaging or summing~\cite{xu2018powerful,duvenaud2015convolutional,defferrard2016convolutional} ignore the distribution information that is crucial to learning graph representation for the power system TSA task. We thus proposed the \emph{distribution-aware learning} module instead of averaging or summation operations to extract more powerful graph representations.


\subsection{Distribution-aware learning}
In statistics, Gaussian distribution is a very important term and is generally used to represent real-valued random variables~\cite{li2016local, chen2020covariance, matsukawa2019hierarchical}. For the design of the final representation, we aim to use a vector representation to outline the distribution information of the Gaussians. Given the node embedding features $\bm{H}={[h_1,...,h_n]}^T \in \mathbb{R}^{n  \times f}$ via GNNs, we design the final graph representation by considering the mean and covariance components of Gaussian, \emph{i.e.}, $\mu\in \mathbb{R}^{f}$ and $\Sigma \in \mathbb{R}^{f \times f}$. The distribution-aware learning (DAL) module for graph representation learning  can be defined as:
\begin{equation}
\label{finalvector}
\begin{split}
\bm{z} =\varphi(\mathcal{N}(\bm{\mu},\bm{\Sigma})) = \bm{\Sigma \mu},
\end{split}
\end{equation}
where $\bm{z}\in \mathbb{R}^{f} $ is the $f$-dimensional resulting graph representation vector. The DAL module can outline the distribution information as presented in the following proposition.


\textbf{Proposition 1.}\; 
\emph{In the Gaussian setting}, the \emph{distribution-aware learning} module $\varphi(\mathcal{N}(\bm{\mu},\bm{\Sigma}) = \bm{\Sigma\mu}\in \mathbb{R}^{f}$ can outline the distribution information of Gaussians.  

\textbf{\textit{Analysis.}}\; From the perspective of mean vector reconstruction in the space spanned by the eigen-vectors of covariance matrix, the Eq. (\ref{finalvector}) can be rewritten as :
\begin{equation}
\begin{split}
\bm{\Sigma \mu} =  \bm{U} \bm{\Lambda} \bm{U}^T  \bm{\mu},
\end{split}
\end{equation}
where the diagonal matrix $\bm{\Lambda}={\rm diag}(\lambda_1,\lambda_2,...,\lambda_f)$ consists of the ordered eigen-values of the covariance matrix, and orthogonal matrix $\bm{U}=[\bm{u}_1,\bm{u}_2,...,\bm{u}_f]$ consists of the normalized eigen-vectors corresponding their eigen-values. In a new linear space spanned by $\bm{U}$, mean vector can be can be represented as $\hat{\bm{\mu}} = {\sum}_{i=1}^{f} {\alpha}_i \bm{u_i} = \bm{U} \bm{\alpha}$,  where $\bm{\alpha} = [\alpha_1,...,\alpha_f]^T \in \mathbb{R}^{f}$ is the coefficient vector related to the basis vectors $[\bm{u}_1,\bm{u}_2,...,\bm{u}_f]$, and $\bm{\mu} = \bm{\hat{\mu}} + \Delta \bm{\mu}$. Note that, as $\Delta \bm{\mu}$ is perpendicular to the space spanned by $\bm{U}$,  $\Delta \bm{\mu}^T \bm{\Sigma}$ will always be zero~\cite{li2019distribution}. Then, the resulting mean vector can be represented as: 
\begin{equation}
\begin{split}
\bm{\Sigma \mu} =  \bm{U} \bm{\Lambda} \bm{U}^T  \bm{U} \bm{\alpha} =  \sum_{i=1}^{f}\lambda_i \alpha_i \bm{u_i} ,
\end{split}
\end{equation}
where the resulting mean vector is the weighted version of $\bm{\hat{\mu}}$, and weights are eigen-values. 
Considering the largest eigenvalue $\lambda_1$, its corresponding eigen-vector $\bm{u_1}$ reflects the direction of maximum variance and represents the main distribution direction of the data. Thus, DAL vector can capture the principal component of the data distribution information by using the eigen-values as weights. Thus, the DAL vector not only contains the information of the mean component but also indicates the distribution information of Gaussians, which is a significant support to our methods.

Furthermore, we show that the DAL module naturally satisfies the two requirements to serve as graph pooling. \emph{Firstly}, pooling method should be able to take $H$ with a variable number of rows as the inputs and produce fixed-sized outputs. \emph{Secondly}, the pooling method should output the exacted representation when the order of rows of $H$ changes.

Now, we give the theoretical proof to check that our proposed DAL module meets the two requirements above. The mean vector $\mu=\frac{1}{n}{\Sigma}_{k=1}^n h_k \in \mathbb{R}^{f}$ obviously meets the two requirements above. The covariance matrix $\Sigma = \frac{1}{n}{\Sigma}_{k=1}^{n} {(h_k-\mu)}^{T}(h_k-\mu) \in \mathbb{R}^{f \times f} = \tilde{H}^T \tilde{H}$, where $\tilde{H}$ is the mean centered feature matrix, can satisfy the two requirements above as presented in the proposition 2 and proposition 3. 

\textbf{Proposition 2.}\; The covariance operation always outputs an $f\times f$ matrix for $H \in \mathbb{R}^{n\times f}$, regardless of the value of $n$.

\textbf{\textit{Proof.}} The result is obvious since the dimension of $H^TH$ does not depend on $n$.

\textbf{Proposition 3.}\;  The covariance matrix is invariant to permutation so that it outputs the same matrix when the order of rows of the feature matrix changes.

\textbf{\textit{Proof.}} Consider $H_1=PH_2$, where $P$ is a permutation matrix. Note that we have $P^TP=I$ for any permutation matrix. Therefore, it is easy to derive

\begin{equation}
\begin{split}
{\rm COV}(H_1) &=  \tilde{H}_1^T\tilde{H}_1  \\
&={(P\tilde{H}_2)}^T{P\tilde{H}_2}  \\
&=\tilde{H}_2^TP^TP\tilde{H}_2\\
&=\tilde{H}_2^T\tilde{H}_2={\rm COV}(H_2).
\end{split}
\end{equation}

According to the above proofs, the mean and covariance satisfy the two requirements to serve as graph pooling. Our proposed DAL module naturally satisfies the two requirements above.

\section{Case Study}
This section studies our proposed \emph{distribution-aware learning} module on two typical systems with different scales, the IEEE 39-bus test system and the Polish 2383-bus power system. MATPOWER~\cite{zimmerman1997matpower} and Power System Analysis Toolbox (PSAT)~\cite{milano2005open} are used for data generation.

\subsection{Graph Representation applied in TSA}

A DAL module embedded graph neural network contains two stages, including \textit{structure-aware learning} and \textit{distribution-aware learning}. Our proposed framework for the TSA task has been shown in Fig.~\ref{fig:framework}. Firstly, we model the power system data as graph structure data by retaining active and reactive power on bus nodes, as illustrated in Fig.~\ref{fig:powersystem_2}. Then, we employ the existing GNN model as the structure-aware learning module to learn the representation vector for the bus nodes. In terms of the learned node feature, we apply our designed DAL module to outline the distribution information among the bus nodes to learn the informative graph representation for describing the status of the power system. Finally, the resulting representation can be treated as the input of the commonly used classification models to achieve the TSA task. Next, we will introduce how to perform this framework for online TSA.

\begin{table}
\vspace{-0.0cm}  
\caption{Stable/unstable samples for two different power system datasets}
\label{sample_statistics}
\renewcommand{\arraystretch}{1.1} 
\normalsize 
\centering
\begin{tabular}{lccccc}
\toprule
          &Stable  &Unstable  &Total     \\
\midrule
IEEE 39-bus		
            &7090	&2910		&10000   		\\
Polish 2383-bus
            &2010	&990		&3000   		\\
\toprule
\end{tabular}
\vspace{-0.4cm}  
\end{table}

\begin{table}
\vspace{0.1cm}  
\caption{Comparison results between our proposed methods with existing benchmark data-driven TSA methods on IEEE 39-bus system. The best model is highlighted with boldface.}
\label{comparison_39}
\renewcommand{\arraystretch}{1.1} 
\normalsize 
\centering
\begin{tabular}{lccccc}
\toprule
Methods          &Acc.  &F1.  &TNR.  &TPR.   \\
\midrule
LR	
            &93.7$\pm$0.6	&95.6$\pm$0.4	&85.2$\pm$1.6   &97.2$\pm$0.5		\\
SVM
            &86.2$\pm$0.4	&91.0$\pm$0.2	&86.1$\pm$1.5 	&98.5$\pm$0.4		\\
LDA		
            &95.2$\pm$0.6	&96.6$\pm$0.4	&86.5$\pm$1.8 	&98.7$\pm$0.3		\\
RF	
            &98.5$\pm$0.4	&98.9$\pm$0.7	&96.7$\pm$1.0 	&99.2$\pm$0.3		\\
XGB	
            &98.7$\pm$0.2	&99.1$\pm$0.1	&97.4$\pm$0.4 	&99.2$\pm$0.2		\\
ANN		
            &98.5$\pm$0.4	&98.9$\pm$0.3	&97.0$\pm$1.0 	&99.0$\pm$0.4		\\
\midrule
GNN
&98.9$\pm$0.3	&99.2$\pm$0.2	&97.5$\pm$1.1 	&99.5$\pm$0.3		\\
Proposed		
		&\textbf{99.2$\pm$0.2}	&\textbf{99.4$\pm$0.2}	&\textbf{98.4$\pm$0.9} &{\textbf{99.6$\pm$0.2}}	 \\		
\toprule
\end{tabular}
\vspace{-0.3cm}  
\end{table}

\section{THE APPLICATION FOR ONLINE TRANSIENT STABILITY ASSESSMENT}

This section performs our framework to online TSA to reduce the average simulation time. The proposed method aims to terminate TDS earlier without losing the accuracy. The online assessment flowchart as shown in Fig.~\ref{flowchart} can be divided into three stages, dataset construction, offline training and online assessment.

\textbf{Dataset construction.} Data-driven methods have been widely used in real applications and require adequate data for training the intelligence model. Therefore,  a large number of transient samples of the power system are vital to the TSA task. The data can be from historical operating records or TDS simulations on different contingencies. However, It is challenging to save and obtain historical data of the actual power grid at the time of disturbance. In this paper, we use the PSAT toolbox to simulate the running status of the power system. We first model the power system to graph structure data by considering the bus nodes' connection information. Then, we define this data-driven TSA problem as a two-class classification problem. 
The label information needs to be clear via Eq. (\ref{labelstable}) for these obtained graph structure data. 
Finally, we need to save these data to construct the TSA dataset for training model.

\textbf{Offline training.} Revisiting the existing deep learning methods, we need to learn the representation of each data to describe the status of the power system. Here, we apply the proposed \emph{distribution-aware} graph representation learning method into the domain of the power system to describe its status. 
We input the processed data into the model and optimize the parameter set. Then, the designed distribution-aware learning module with the learned knowledge can be applied for online assessment.

\textbf{Online assessment.} At the online stage, the proposed method is triggered when there is a physical fault occurring in the system. Following the fault, the transient stability of the system is assessed in a self-adaptive way based on the proposed method.  With the data-driven based methods in this paper, we need to evaluate the credibility of TSA results. As shown in Fig. \ref{flowchart}, if the value of $|S_0-S_1|$ is less than a set threshold, we think that the model is inaccurate in determining the sample and selecting the TDS for analyzing the status of the power system. Suppose the absolute value of the difference between the two is not less than this threshold. In that case, we believe that the model is sufficiently confident in the given predictions and assess the status of the power system by comparing the value of $S_0$ and $S_1$.

\textbf{IEEE 39-bus System.}
The IEEE 39-bus power system has 39 buses, 10 generators, 19 loads and 46 transmission lines. All generators in this system use a 4th-order model, with a 4th-order excitation system~\cite{milano2005open}. The Appendix in~\cite{pai2012energy} gave steady-state and transient parameters of this system, which is essential for transient stability analysis.
We follow~\cite{yan2019fast,han2021imbalanced} to adopt the following principles for generating unstable data.
\begin{itemize}
    \item We randomly change active and reactive power on all loads within $80\%$ to $120\%$ of the primary load level.

    \item We use the MATPOWER toolkit to compute the optimal power flow for the next time domain simulation.

    \item We consider a three-phase to ground fault for our task and clear it after a random time within 1/60 to 1/6 seconds.

    \item We manually label the generated sample after 10 second TDS simulation using PSAT.
\end{itemize}
As shown in Table~\ref{sample_statistics}, we generated a total of 10000 samples, including 7090 stable (positive) samples and 2910 (negative) samples, to train the intelligence model.

\textbf{Polish 2383-bus System.}
The Polish 2383-bus system contains 2383 buses, 327 generators, 1561 loads and 2896 transmission lines. All generators use the 6th-order model, with 4th-order excitation and 1st-order governor control systems. In addition, some generators include a 4rd-order power system stabilizer. The means for generating samples is the same as the IEEE 39-bus system. As shown in Table~\ref{sample_statistics}, there are  2010 stable samples and 990 unstable samples generated for training data-driven model.

\begin{table}
\vspace{0.1cm}  
\caption{Comparison results between our proposed methods with existing benchmark data-driven TSA methods on Polish 2383-bus system. The best model is highlighted with boldface.}
\label{comparison_2383}
\renewcommand{\arraystretch}{1.1} 
\normalsize 
\centering

\begin{tabular}{lccccc}
\toprule
Methods          &Acc.  &F1.  &TNR.  &TPR.   \\
\midrule
LR		
            &98.4$\pm$0.5	&98.8$\pm$0.4	&98.1$\pm$1.2   &98.6$\pm$0.4		\\
SVM	
            &92.9$\pm$1.0	&94.9$\pm$0.7	&80.5$\pm$2.9 	&99.0$\pm$0.4		\\
LDA		
            &88.5$\pm$2.1	&90.8$\pm$1.8	&85.1$\pm$2.9 	&95.7$\pm$1.9		\\
RF	
            &95.8$\pm$1.3	&96.9$\pm$0.9	&92.1$\pm$2.3 	&97.7$\pm$1.3		\\
XGB	
            &98.6$\pm$0.9	&99.0$\pm$0.4	&98.5$\pm$1.0 	&98.8$\pm$0.4		\\
ANN		
            &98.7$\pm$0.6	&99.1$\pm$0.3	&98.7$\pm$1.1 	&98.9$\pm$0.6		\\

\midrule
GNN
		&99.0$\pm$0.6	&99.2$\pm$0.4	&98.9$\pm$1.2 	&99.1$\pm$0.8		\\

Proposed		
 		&\textbf{99.2$\pm$0.5}	&\textbf{99.4$\pm$0.4}	&\textbf{99.2$\pm$0.8} 	&\textbf{99.3$\pm$0.7}		\\
\toprule
\end{tabular}
\vspace{-0.3cm}  
\end{table}

\begin{table}
\vspace{-0.cm}  
\caption{Ablation study about network layer setting on IEEE 39-bus system and Polish 2383-bus system. Our
method achieves satisfactory performance while setting the layer as 4.}
\label{results_layers_39}
\renewcommand{\arraystretch}{1.08} 
\normalsize 
\centering

\begin{tabular}{l m{12mm}<{\centering} m{12mm}<{\centering} m{12mm}<{\centering} m{12mm}<{\centering} m{12mm}<{\centering}}
\toprule
&          &Acc.  &F1.  &TNR.  &TPR.   \\
\midrule
\multirow{4}{*}{\textbf{\rotatebox{90}{39-bus}}}
&Layer-2		
            &98.8$\pm$0.3	&99.1$\pm$0.2	&97.3$\pm$0.8   &99.4$\pm$0.3		\\
&Layer-3
            &99.1$\pm$0.2	&99.4$\pm$0.2	&98.0$\pm$0.5 	&99.6$\pm$0.3		\\
&Layer-4	
            &99.2$\pm$0.2	&99.4$\pm$0.2	&98.4$\pm$0.9 	&99.6$\pm$0.2		\\
&Layer-5	
            &99.2$\pm$0.2	&99.4$\pm$0.2	&98.1$\pm$0.7 	&99.6$\pm$0.3		\\

\midrule
\multirow{4}{*}{\textbf{\rotatebox{90}{2383-bus}}}
&Layer-2		
            &97.9$\pm$0.7	&98.4$\pm$0.5	&97.9$\pm$1.7   &97.9$\pm$1.1		\\
&Layer-3	
            &98.9$\pm$0.7	&99.1$\pm$0.5	&98.9$\pm$1.3 	&98.8$\pm$0.9		\\
&Layer-4	
            &99.2$\pm$0.5	&99.4$\pm$0.4	&99.2$\pm$0.8 	&99.3$\pm$0.7		\\
&Layer-5	
            &99.2$\pm$0.5	&99.4$\pm$0.5	&98.5$\pm$1.0 	&99.6$\pm$0.2		\\

\bottomrule
\end{tabular}
\vspace{-0.2cm}  
\end{table}

\begin{figure*}
  \vspace{-0.4cm}  
  \setlength{\abovecaptionskip}{0.2cm} 
\centering
\hspace{-4mm}
\subfigure[Fold 1]{\includegraphics[width=3.8cm]{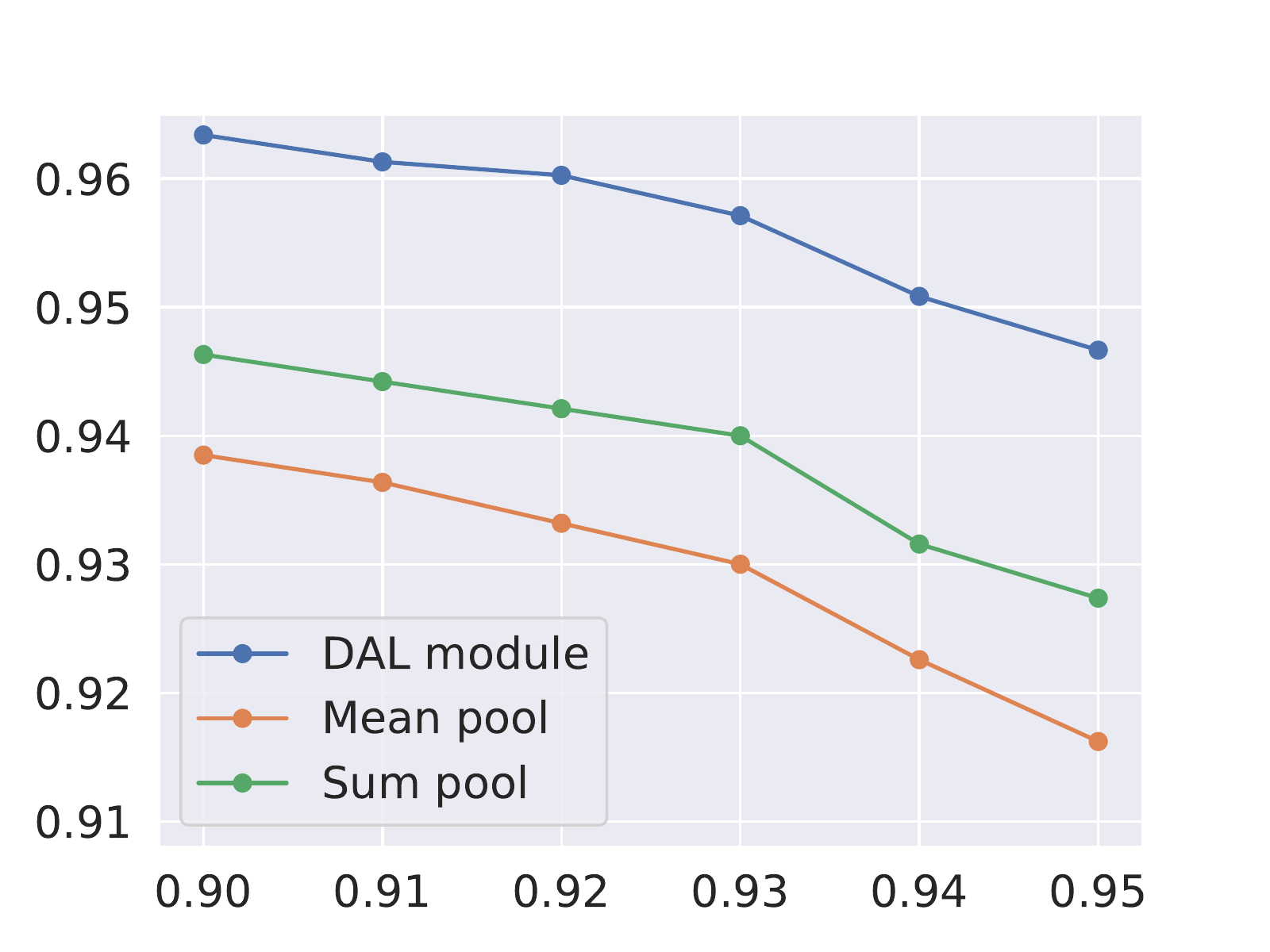}}
\hspace{-4mm}
\subfigure[Fold 2]{\includegraphics[width=3.8cm]{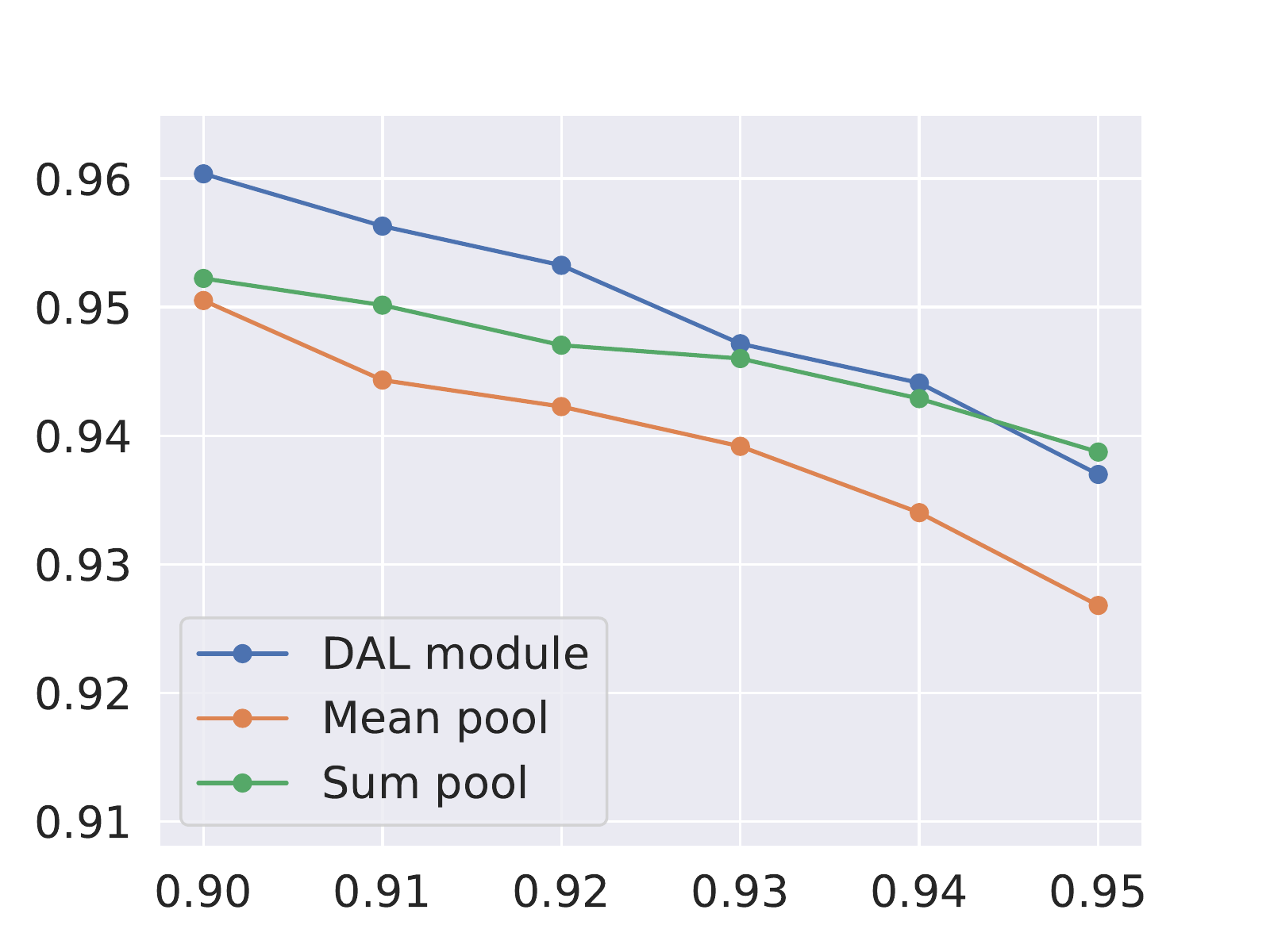}}
\hspace{-4mm}
\subfigure[Fold 3]{\includegraphics[width=3.8cm]{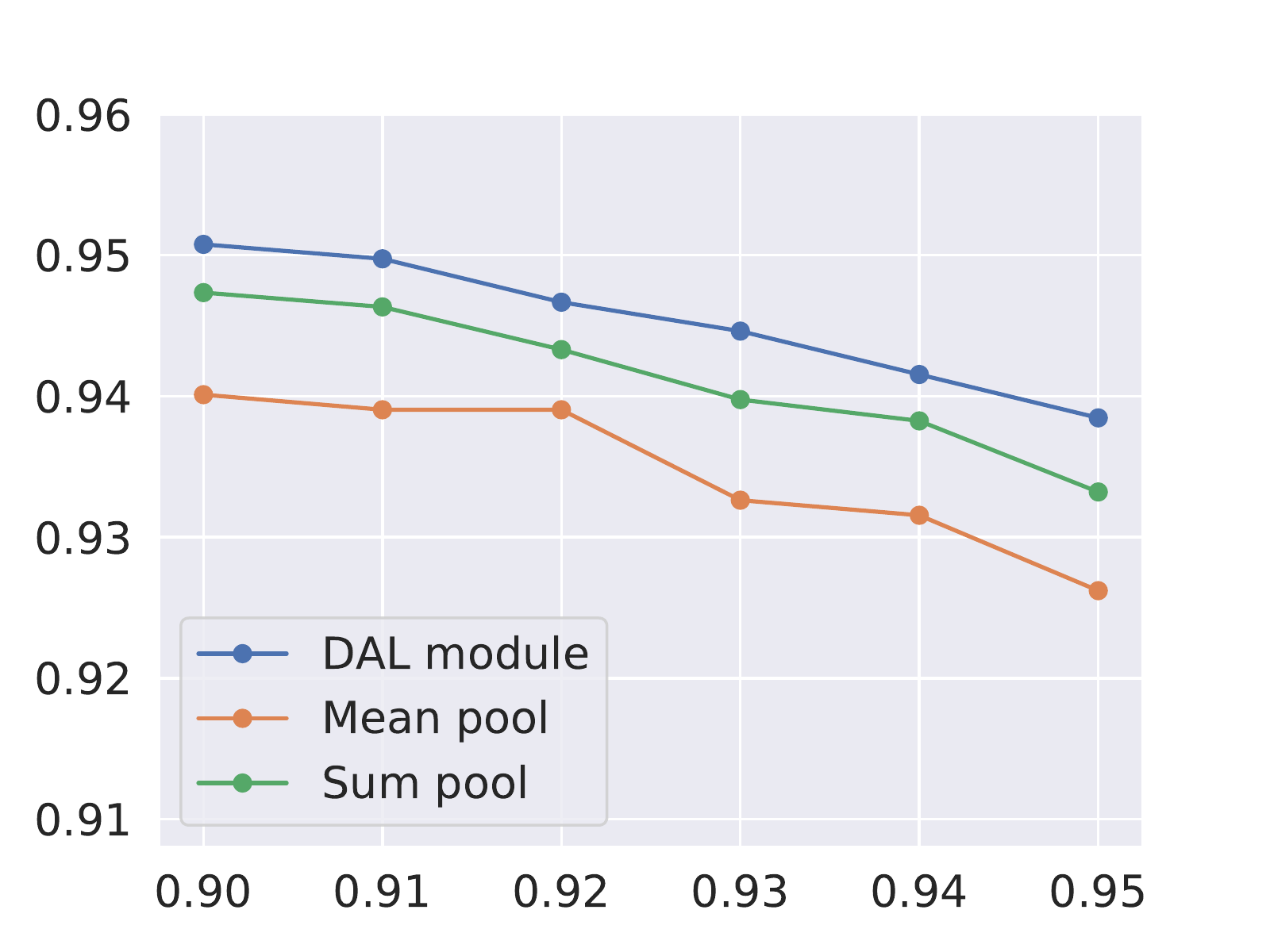}}
\hspace{-4mm}
\subfigure[Fold 4]{\includegraphics[width=3.8cm]{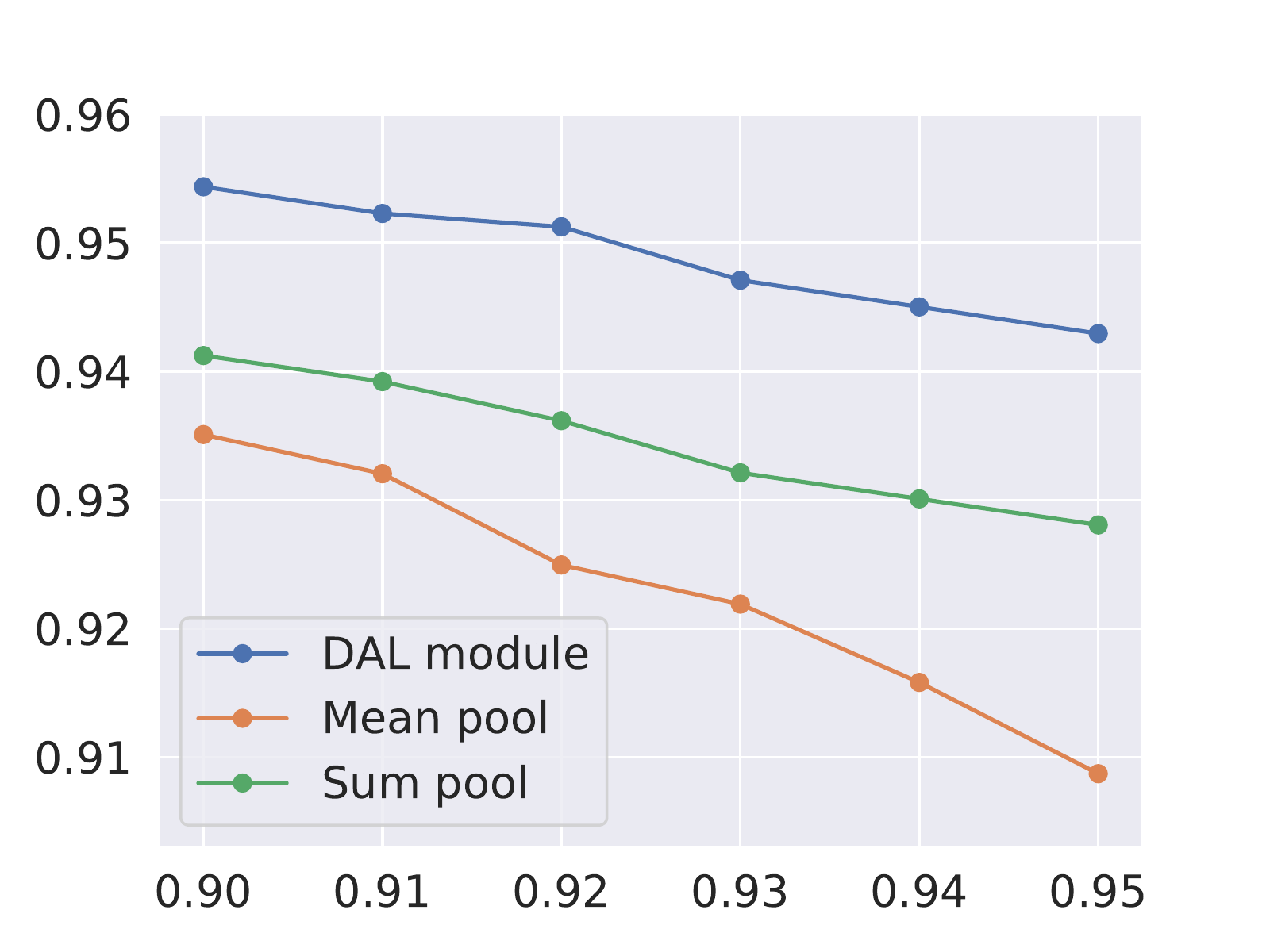}}
\hspace{-4mm}
\subfigure[Fold 5]{\includegraphics[width=3.8cm]{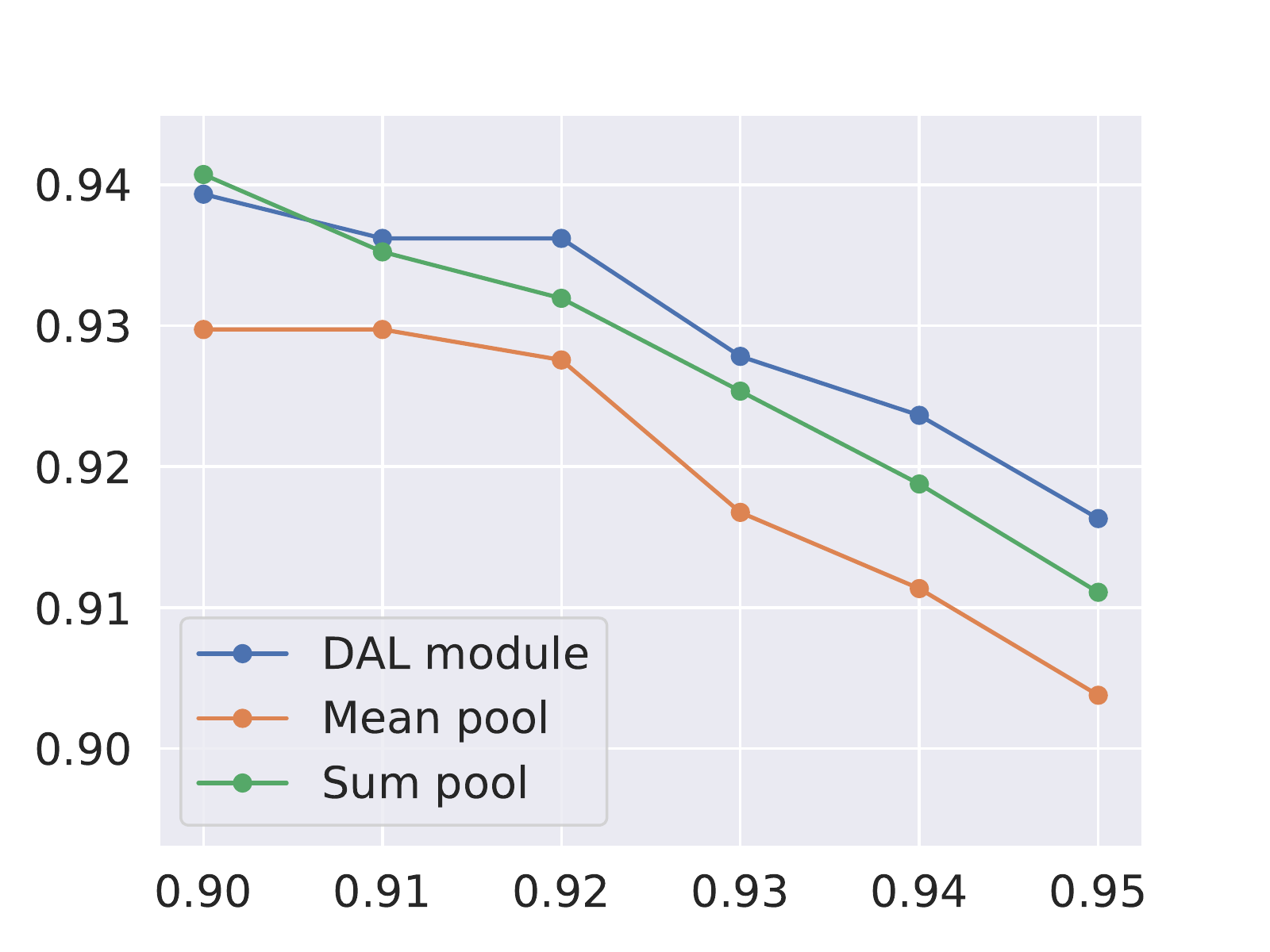}}
\hspace{-4mm}

\vspace{-2mm}

\hspace{-4mm}
\subfigure[Fold 6]{\includegraphics[width=3.8cm]{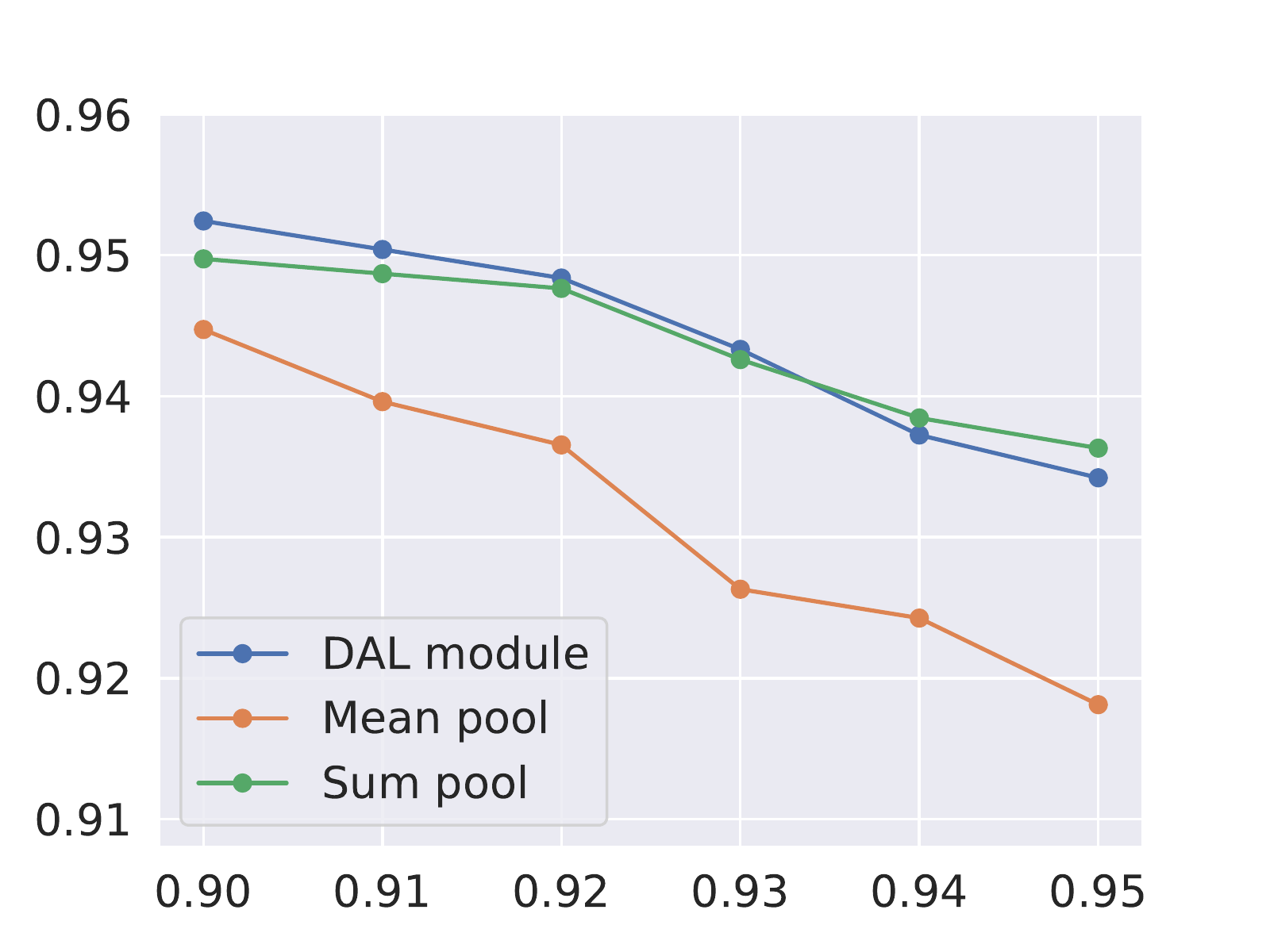}}
\hspace{-4mm}
\subfigure[Fold 7]{\includegraphics[width=3.8cm]{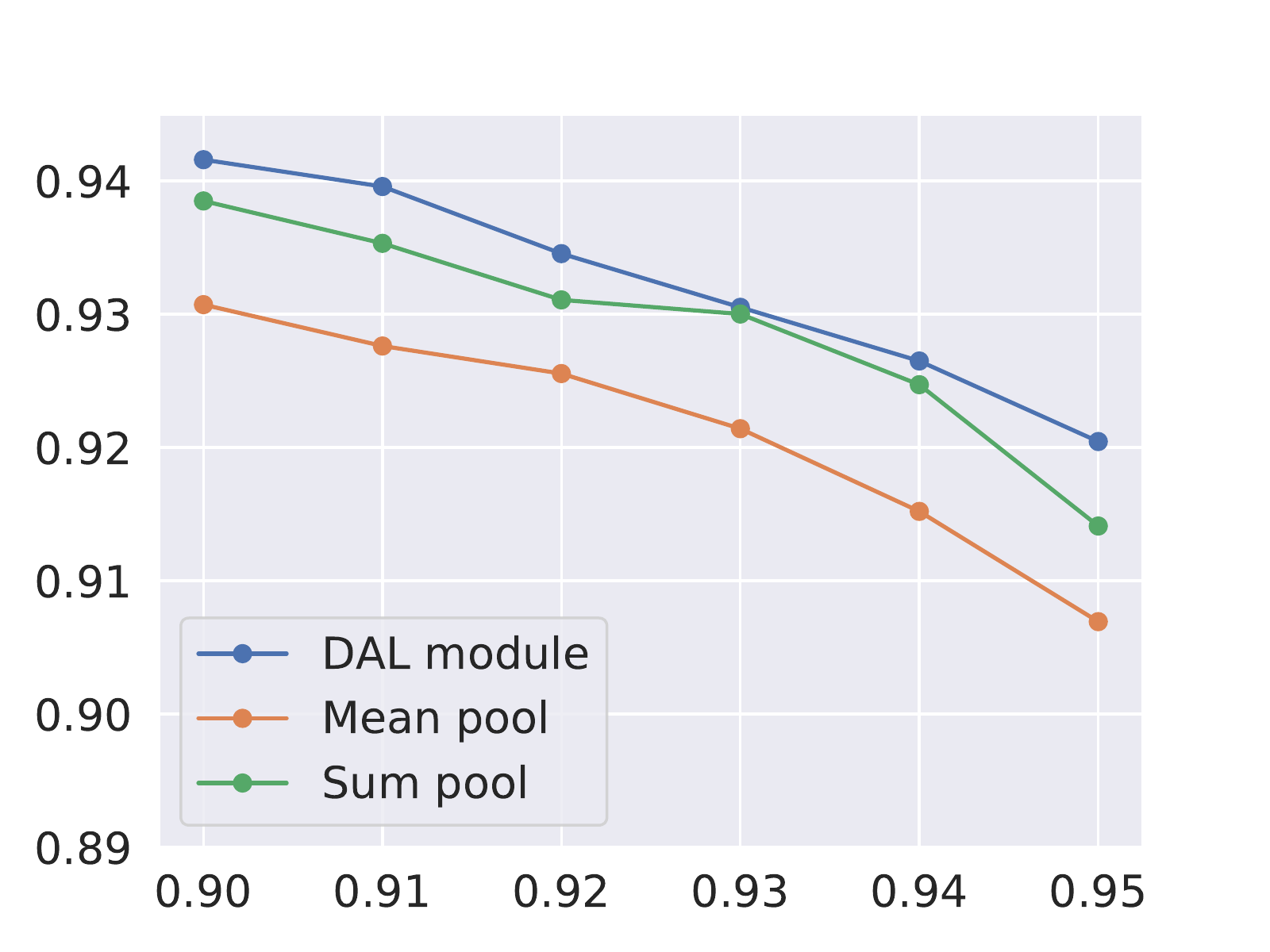}}
\hspace{-4mm}
\subfigure[Fold 8]{\includegraphics[width=3.8cm]{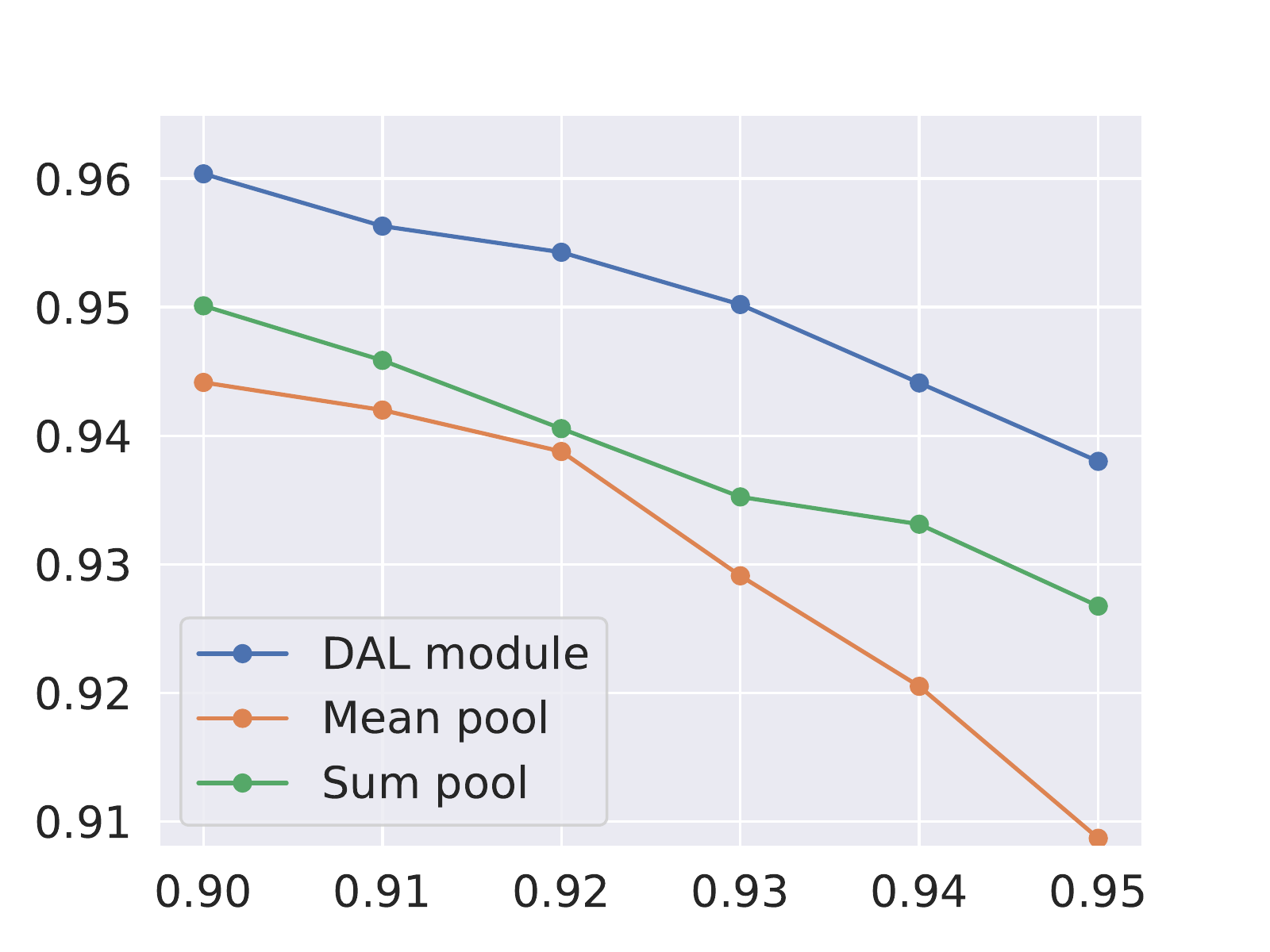}}
\hspace{-4mm}
\subfigure[Fold 9]{\includegraphics[width=3.8cm]{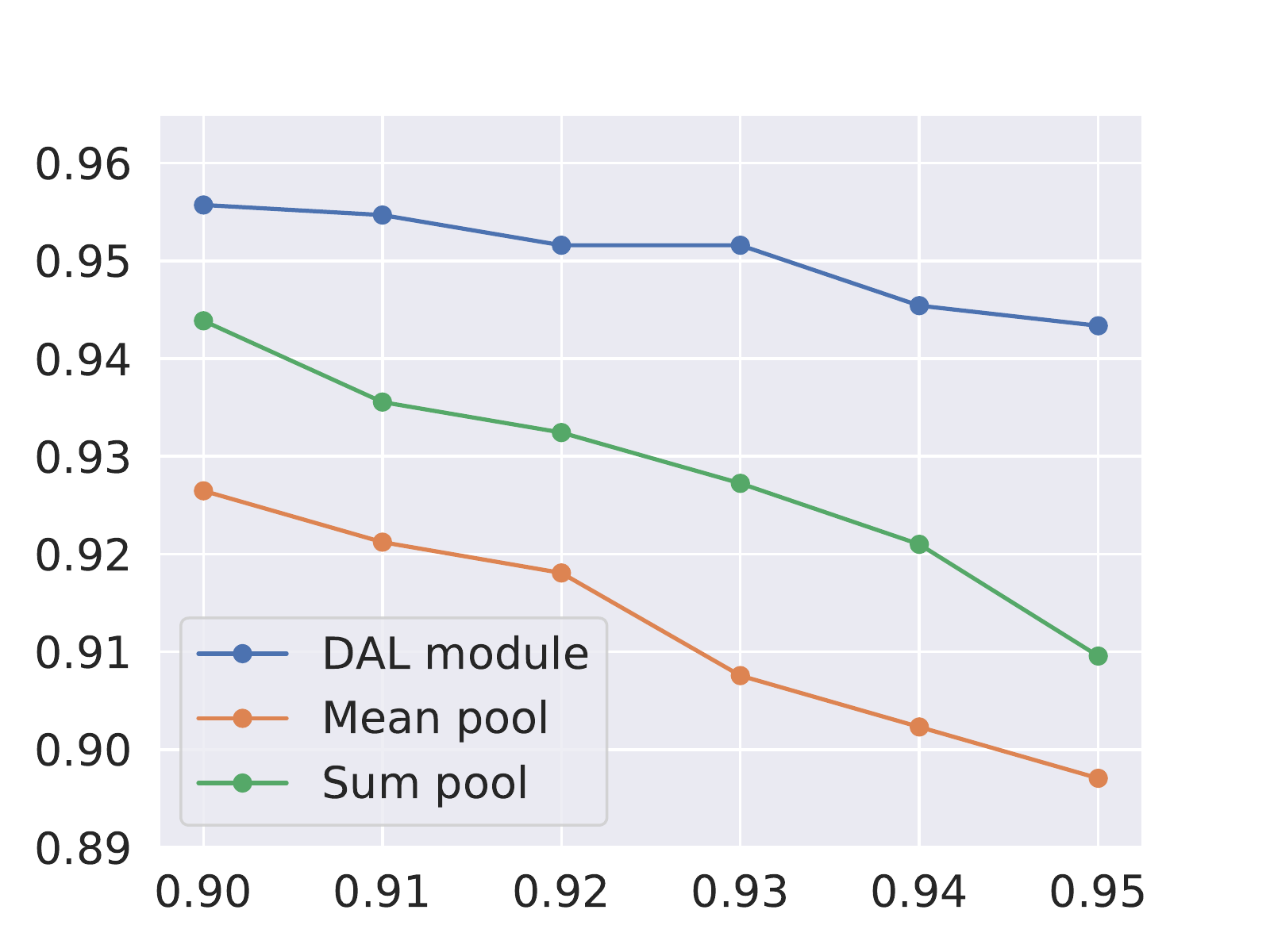}}
\hspace{-4mm}
\subfigure[Fold 10]{\includegraphics[width=3.8cm]{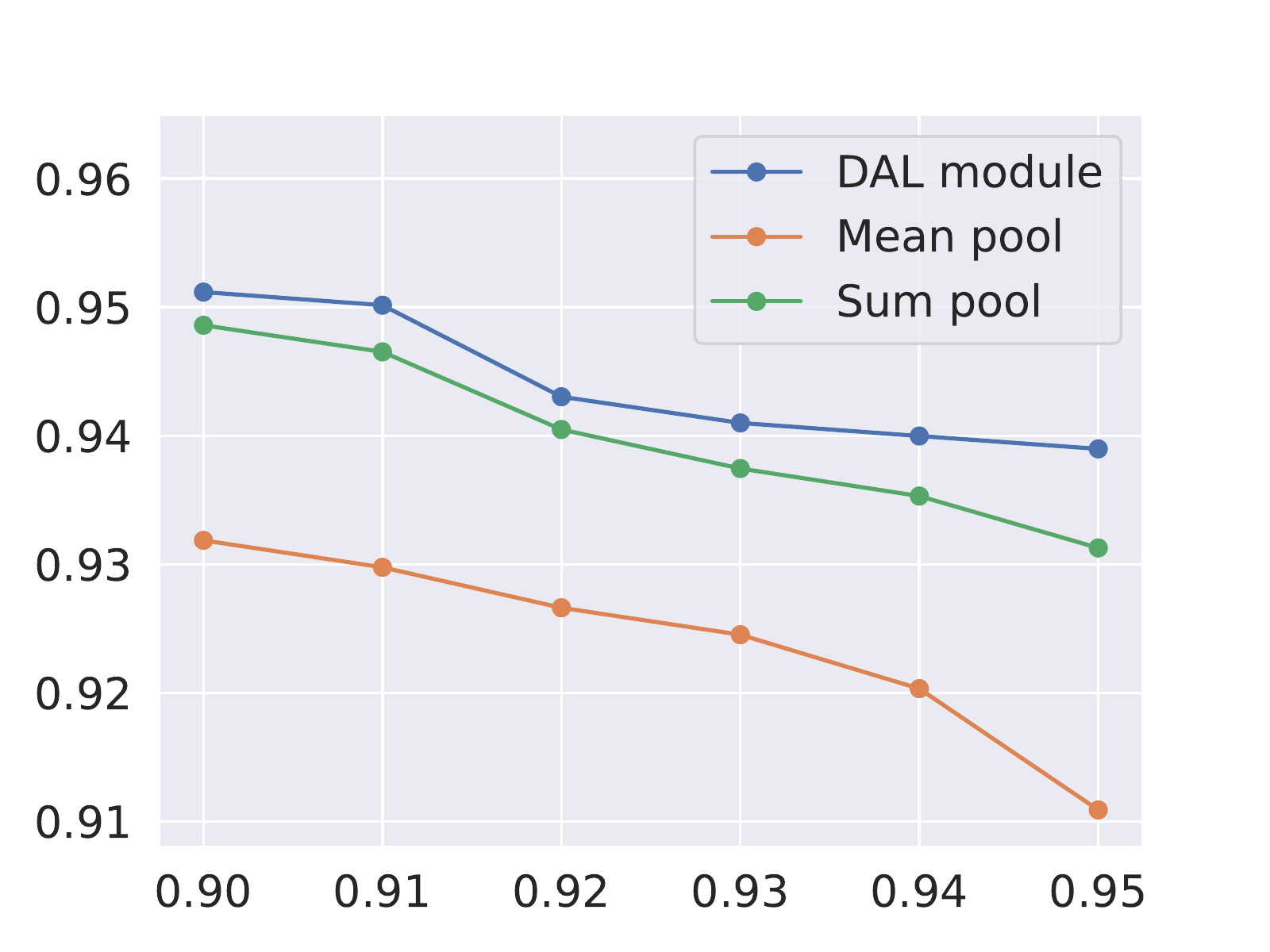}}
\hspace{-4mm}

\caption{
Ablation study about credibility rating of each fold on IEEE 39-bus system. The abscissa refers to the threshold $k$, and ordinate refers to the credibility rating index.
}
\label{threshold_10folds}
  \vspace{-0.4cm}  
\end{figure*}

\begin{figure}
\vspace{-0.2cm}  
  \vspace{-0.2cm}  
  \setlength{\abovecaptionskip}{0.cm} 
  \setlength{\belowcaptionskip}{0.cm} 
\centering
\hspace{-2mm}
\subfigure[Polish 2383-bus]{\includegraphics[width=4.6cm]{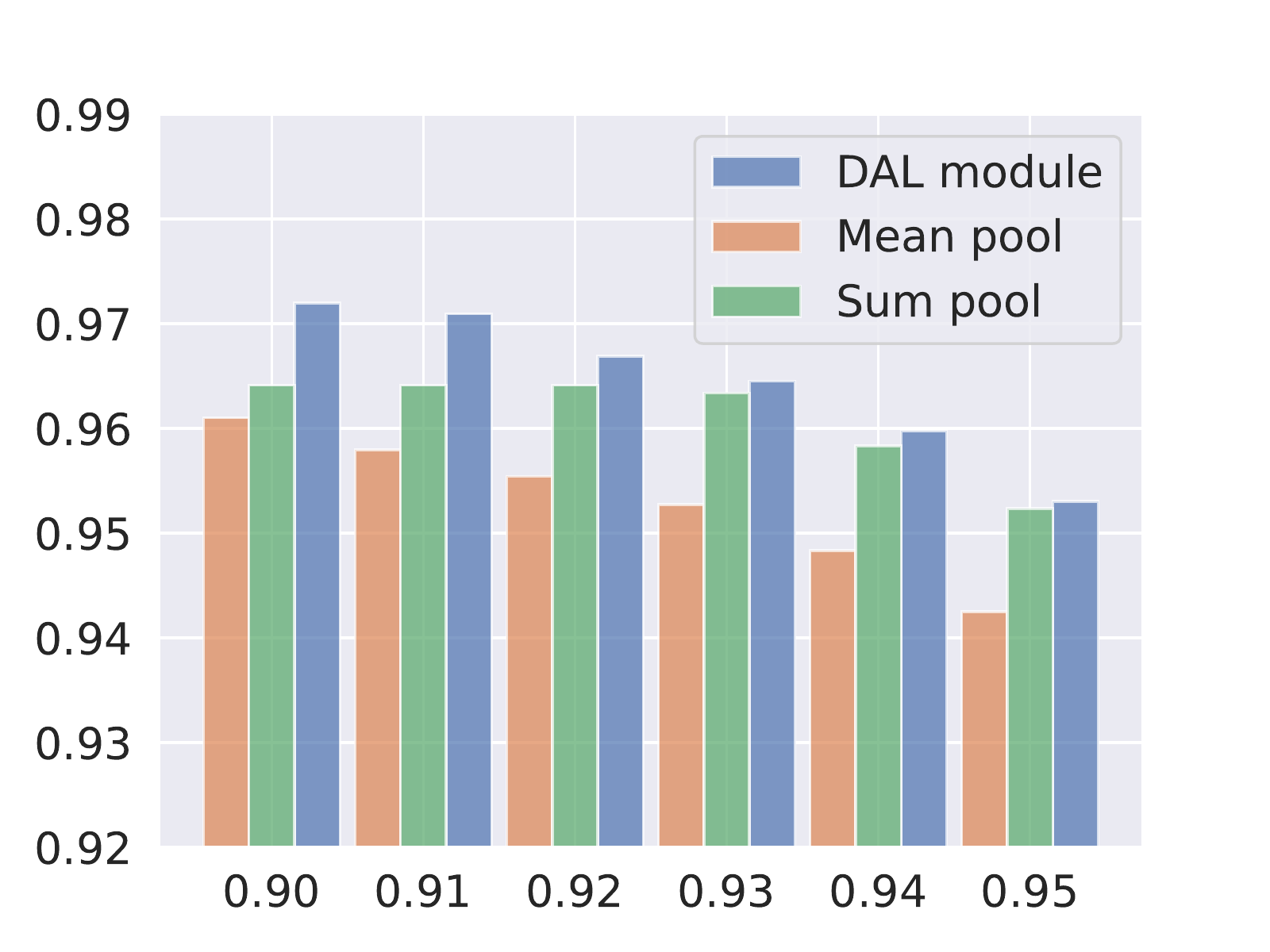}}
\hspace{-4mm}
\subfigure[IEEE 39-bus]{\includegraphics[width=4.6cm]{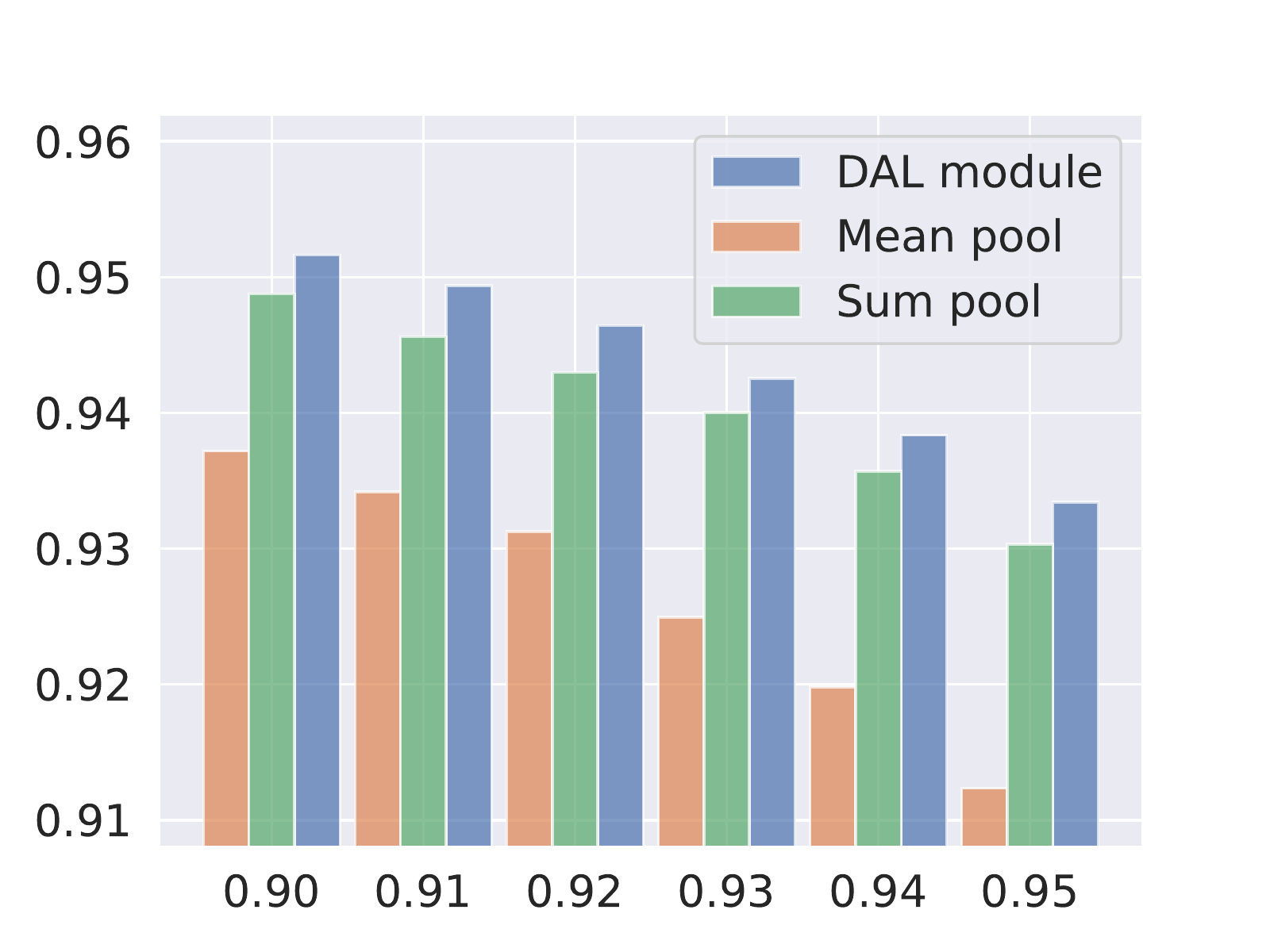}}
\hspace{-4mm}

\caption{
Ablation study about credibility rating on IEEE 39-bus and Polish 2383-bus systems. The abscissa refers to the threshold $k$, and ordinate refers to the credibility rating index.
}
\label{threshold}
\vspace{-0.4cm}  
\end{figure}

\subsection{Experimental evaluation}
To verify the effectiveness of our methods, we perform comparison experiments with several most used data-driven methods, including logistic regression, support vector machine, random forest, XGBoost, and artificial neural network. Furthermore, we use the following four indicators to evaluate the performance among different models, including F1 score, accuracy (Acc), true negative rate (TNR) and true positive rate (TPR).

We perform 10-fold cross-validation suggested in~\cite{xu2018powerful} to calculate the mean values and standard deviations following the above four indicators. The experimental results of different methods on two power system datasets are presented in Tables~\ref{comparison_39} and~\ref{comparison_2383}, respectively. The GNN method in the Table \ref{comparison_39} and \ref{comparison_2383} denotes the GIN network~\cite{xu2018powerful} in our experiments. Our proposed method just replaces the GIN pooling module with other components unchanged, and also adopts the same experimental setting in this paper. According to the comparison results in the Tables, our model achieves the best performance based on the above indicators. Thus, we summarize that GNN based method can achieve better performance because they consider the topological structure of the power system. Furthermore, our proposed DAL module plugged at the end of GNN can further improve the performance of the power system TSA task.

\subsection{Ablation studies}
In addition to the overall recognition rate used for verifying the effectiveness of our method, we also need to consider the credibility according to the value of  threshold $k \leq  |S_0-S_1|$.  The credibility rating means the proportion of correctly classified samples that meet the threshold value in all correctly classified samples.
\begin{equation}
\begin{split}
{\rm CR} = \frac{num({\rm P(sample)} \cap  {\rm T(sample)})}{num({\rm P(sample)})} ,
\end{split}
\end{equation}
where $num$ refers to the number of samples that meet certain requirements, P denotes those samples whose predictions are correct, T denotes those samples that meet the threshold $k$, $\cap$ denotes the intersection of samples. The CR denotes the ratio among the test cases that can be directly applied to online assessment, which is an important indicator to evaluate the performance of the model at the online application stage.

We analyze our proposed pooling module compared with the most used mean and summation pooling for TSA task. As shown in Fig.~\ref{threshold_10folds} and Fig.~\ref{threshold}, we give the details of each fold on the IEEE 39-bus dataset and the mean accuracies of 10-fold on two datasets. According to the above results, our designed module is not only better than the existing methods in terms of the recognition rate, but also better than existing methods in terms of the CR. In addition, we also consider the influence of the number of network layers. As shown in Table \ref{results_layers_39},  our method will achieve satisfactory performance
while setting the layer as 4.

\section{Conclusion}
In this paper, we aim to describe the status of the power system using the graph representation learning method. Motivated by observing the structure of the power system and the distribution of the activate and reactive on the bus nodes, we propose a practical module plugged at the end of existing graph neural networks to learn a more informative graph-level representation.
To embed the Gaussian into a Euclidean space, we propose a distribution-aware module and provide theoretical analysis to support why our method can outline distribution information. Furthermore, we proved that it is reasonable to meet the requirements of the graph pooling
operator. Finally, we evaluate our method on the two classical power systems and show superior performance compared with the other methods for TSA task.

\section*{Acknowledgment}

This work is funded by the National Key R\&D Program of China (Grant No: 2018AAA0101503) and the Science and technology project of SGCC (State Grid Corporation of China): fundamental theory of human-in-the-loop hybrid-augmented intelligence for power grid dispatch and control.

\bibliographystyle{IEEEtran}
\bibliography{ref}

\end{document}